\documentclass[11pt, a4paper]{article}

\usepackage[margin=1in]{geometry} % Standard margins
\usepackage{authblk}              % For author affiliations
\usepackage{times}                % Times New Roman font (standard for papers)
\usepackage[hidelinks]{hyperref}  % Clickable links in PDF

\usepackage{amsmath}
\usepackage{graphicx}
\usepackage{natbib}
\usepackage{amsthm}
\usepackage{color}
\usepackage{footmisc}
\usepackage{url}

\usepackage{tikz}
\usetikzlibrary{shapes.geometric, arrows.meta, positioning, fit, backgrounds}
\usepackage{tabularx}
\usepackage{booktabs}
\usepackage{ragged2e} 

\usepackage{listings}
\usepackage{xcolor}
\title{Calliope: A TTS-based Narrated E-book Creator Ensuring Exact Synchronization, Privacy, and Layout Fidelity}

\author[1,2]{Hugo L. Hammer}
\author[2]{Vajira Thambawita}
\author[2,1]{Pål Halvorsen}

\affil[1]{Oslo Metropolitan University, Oslo, Norway}
\affil[2]{SimulaMet, Oslo, Norway}

\date{} % Leave empty to omit the date, or put \today

\begin{document}

\maketitle

\begin{abstract}
A narrated e-book combines synchronized audio with digital text, highlighting the currently spoken word or sentence during playback. This format supports early literacy and assists individuals with reading challenges, while also allowing general readers to seamlessly switch between reading and listening. With the emergence of natural-sounding neural Text-to-Speech (TTS) technology, several commercial services have been developed to leverage these technology for converting standard text e-books into high-quality narrated e-books. However, no open-source solutions currently exist to perform this task. In this paper, we present \textbf{Calliope}, an open-source framework designed to fill this gap. Our method leverages state-of-the-art open-source TTS to convert a text e-book into a narrated e-book in the EPUB 3 Media Overlay format. The method offers several innovative steps: audio timestamps are captured directly during TTS, ensuring exact synchronization between narration and text highlighting; the publisher’s original typography, styling, and embedded media are strictly preserved; and the entire pipeline operates offline. This offline capability eliminates recurring API costs, mitigates privacy concerns, and avoids copyright compliance issues associated with cloud-based services. The framework currently supports the state-of-the-art open-source TTS systems XTTS-v2 and Chatterbox. A potential alternative approach involves first generating narration via TTS and subsequently synchronizing it with the text using forced alignment. However, while our method ensures exact synchronization, our experiments show that forced alignment introduces drift between the audio and text highlighting significant enough to degrade the reading experience. Source code and usage instructions are available at 
%(blinded since the URL reveals the author(s)). 
\url{https://github.com/hugohammer/TTS-Narrated-Ebook-Creator.git}.
\end{abstract}

\vspace{0.5em}

\section{Introduction}

“Narrated E-book”, also referred to as “Read Aloud” or “Immersive reading”, refers to technology that reads text aloud while highlighting the word or sentence currently being spoken. This feature supports early readers in developing literacy skills and assists individuals with dyslexia or other reading challenges~\cite{jung2025impact}. It also benefits general readers by enabling seamless switching between reading and listening, depending on their preference at any given moment.

Several technologies enable synchronized text and audio playback. A key open-source standard is EPUB 3 Media Overlays, which uses SMIL (Synchronized Multimedia Integration Language) to align audio narration with text content~\cite{epub3spec}. Another major standard is DAISY (Digital Accessible Information System), designed primarily for individuals with print disabilities~\cite{daisystandard}. EPUB 3 Media Overlays modernize and generalize DAISY concepts. For web-based real-time applications, the W3C Web Speech API allows developers to implement text-to-speech (TTS) and speech recognition directly in browsers, supporting text highlighting during speech synthesis~\cite{webspeechapi}.

\begin{figure}[!t]
    \centering
    \includegraphics[width=\linewidth]{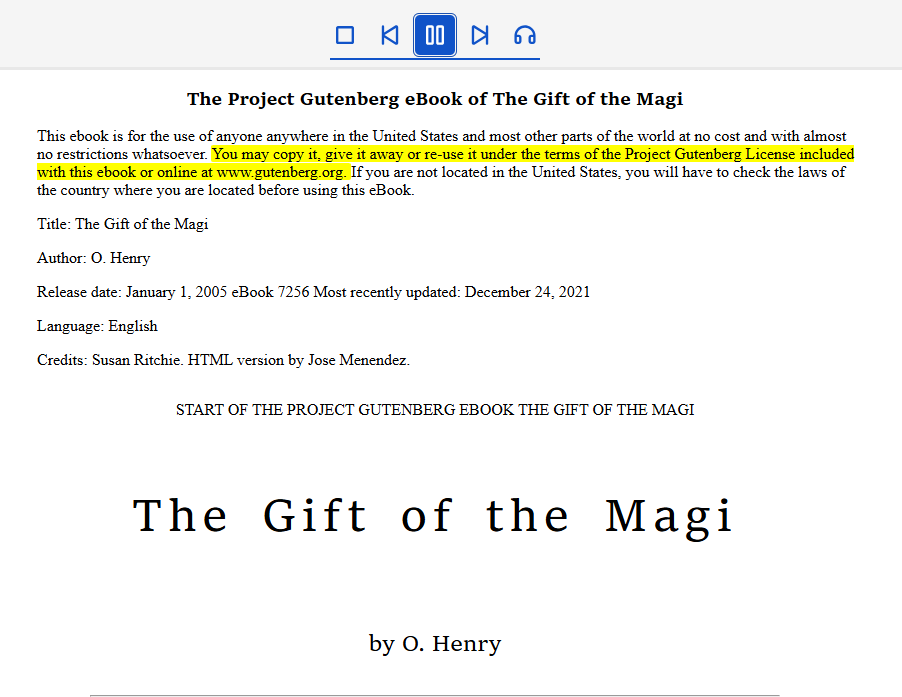}
    \caption{Example of an EPUB 3 Media Overlay file created using the proposed method. The figure demonstrates active text highlighting synchronized with audio playback, while the publisher's original layout and styling are preserved.}
    \label{fig:1}
\end{figure}

Recent years have seen the emergence of natural-sounding neural Text-to-Speech (TTS) technology. Several commercial services, such as ElevenReader and NaturalReader, leverage this technology to convert standard text e-books into high-quality narrated e-books~\cite{naturalreader,elevenreader}. However, currently, no open-source solutions exist to perform this task. To address this gap, we present \textbf{Calliope}, an open-source framework designed to fill this void. Our method converts an e-book in the EPUB format into a narrated e-book in the EPUB 3 Media Overlay format, obtaining audio via state-of-the-art (SOTA) open-source TTS models. The implementation supports both XTTS-v2~\cite{casanova2024xtts} and Chatterbox~\cite{chatterboxtts2025} for audio generation. Calliope incorporates several innovative steps to ensure high quality: audio timestamps are computed directly during the TTS process, ensuring exact synchronization between text highlighting and audio; the pipeline performs surgical modifications of the EPUB container files, preserving the publisher's original typography, styling, and embedded media; and it operates entirely offline, ensuring user privacy and maintaining strict control over copyrighted material. Figure \ref{fig:1} illustrates an example of a resulting EPUB 3 Media Overlay file created using our methodology when played in the Thorium reader\footnote{A short video demonstration is available at \url{https://youtu.be/j15BHY1hh7w}.}\cite{thorium2025}. 
As shown, the text is highlighted (yellow) and synchronized with the audio, while the publisher's original layout and styling remain preserved.

Several related open-source or partly open-source solutions exist to facilitate the creation of narrated e-books. Solutions such as \textit{Storyteller}~\cite{storyteller} and \textit{Syncabook}~\cite{syncabook} function as forced alignment engines, requiring pre-existing audiobook files and using algorithms such as Dynamic Time Warping (DTW) to map the audio stream to the text~\cite{sakoe1978dynamic,Pettarin_aeneas_2016}. Forced alignment tools are prone to synchronization drift, resulting in misalignment between the highlighting and the audio. In some cases, this leads to audio truncation or, depending on the reader implementation, time-stretching artifacts where the audio sounds distorted. In principle, one could generate audio using TTS and then apply forced alignment, but the experiments in this paper show that forced alignment results in drift between audio and text highlighting significant enough to reduce the reading experience. \textit{Epub2tts}~\cite{epub2tts} leverages SOTA TTS models, such as Chatterbox, but does not create narrated e-books, only audiobooks. Finally, \textit{Audible-epub3-maker}~\cite{audibleepub3maker} generates synchronized EPUB 3 Media Overlays but relies on commercial cloud APIs. This reliance introduces recurring costs and necessitates the uploading of sensitive or copyrighted material to third-party servers. Furthermore, existing open-source pipelines employ a ``reconstruction" strategy that strips the original CSS and layout, degrading the visual fidelity of the resulting narrated e-book. Table \ref{tab:2} provides an overview of the main properties of our proposed method (first row) compared to existing open-source solutions.
\begin{table}[ht]
\centering
\caption{Comparison with existing open-source solutions.}
\label{tab:2}
\small 
\setlength{\tabcolsep}{2pt}
% CHANGE 1: Use 'tabular' instead of 'tabularx'
% CHANGE 2: Remove the width argument {\columnwidth}
\begin{tabular}{@{} 
    p{3.0cm}   % Method - adjusted slightly for 2-col fit
    p{2.6cm}   % Sync Strategy
    p{1.7cm}   % Privacy
    p{2.8cm}   % Layout
    p{1.7cm}   % Cost
    p{2.2cm}   % Sync Acc
    @{}}
\toprule
\textbf{Method} & \textbf{Sync} \newline \textbf{Strategy} & \textbf{Privacy} & \textbf{Layout} & \textbf{Cost} & \textbf{Sync.} \newline \textbf{Accuracy} \\ \midrule
\textbf{Calliope} & Deterministic & Local & Preserved & Free & Exact \\ \addlinespace
\textit{Syncabook}~\cite{syncabook} & Forced \newline alignment & Local & Reconstructed & Free & Variable \\ \addlinespace
\textit{Storyteller}~\cite{storyteller} & Transcribe \newline + alignment & Local & Reconstructed & Free & Variable \\ \addlinespace
\textit{Epub2tts}~\cite{epub2tts} & Audio only & Local & N/A & Free & N/A \\ \addlinespace
\textit{Audible-maker}~\cite{audibleepub3maker} & Cloud TTS & Cloud & Reconstructed & Paid & Unknown \\ \bottomrule
\end{tabular}
\end{table}

The main contributions of this paper are:
\begin{itemize}
\item We present an open-source, offline solution for creating EPUB 3 files with Media Overlays using TTS.
\item We employ two state-of-the-art open-source TTS methods: XTTS-v2~\cite{casanova2024xtts} and Chatterbox~\cite{chatterboxtts2025}.
\item The offline architecture preserves user privacy and mitigates copyright risks associated with cloud-based services.
\item The solution performs surgical modifications to the source file, ensuring preservation of the publisher's original layout and styling.
\item Synchronization timings are computed directly during TTS generation, guaranteeing perfect alignment between text highlighting and audio.
\item We introduce a recursive text segmentation algorithm that mitigates the fixed context-window constraints of Transformer-based TTS models, enabling seamless processing of complex sentence structures without truncation or hallucination.
%\item The solution injects a dynamic, high-specificity CSS layer that adapts highlighting contrast based on user system preferences (e.g., Dark Mode).
\end{itemize}

\section{The EPUB 3 Standard}

%To understand our proposed methodology for creating EPUB 3 Media Overlays using TTS, this section first describes the structure of the EPUB 3 standard. 
An EPUB publication is fundamentally a zipped container that follows the Open Container Format (OCF)~\cite{epub3spec}. Three key components within the EPUB container are:

\begin{enumerate}
\item \textbf{Package Document (OPF):} This XML file serves as the central “brain” of the publication. It contains the \textit{Metadata} (bibliographic information), the \textit{Manifest} (a complete inventory of all files in the archive), and the \textit{Spine} (the linear reading order of the XHTML content documents).
\item \textbf{Content Documents (XHTML):} These files contain the narrative text and structural markup (HTML5). They also reference auxiliary resources such as images and stylesheets.
\item \textbf{Style Sheets (CSS):} These define the visual presentation (fonts, margins, colors) and are linked via standard HTML \texttt{<link>} tags within the XHTML headers.
\end{enumerate}

To transform a static e-book into a synchronized audiobook, EPUB 3 introduces \textit{Media Overlays}~\cite{epub3spec}. This feature establishes a triangulation of references between the text, the audio, and a synchronization map, as described below. 
%Our pipeline automates the generation of these components as described in the sections below.
%and performs precise injections of cross-references into the OPF manifest. 
%The following sections describe this triangulation between text, audio, and synchronization in detail.

\subsection{The Content Document (XHTML)}
The text content is segmented, with each synchronization unit (sentence or word) wrapped in a \texttt{<span>} element containing a unique identifier (\texttt{id}). This identifier serves as the anchor for the synchronization logic:
\begin{lstlisting}[language=xml]
<?xml version="1.0" encoding="UTF-8"?>
<html xmlns="http://www.w3.org/1999/xhtml" 
      xmlns:epub="http://www.idpf.org/2007/ops" 
      lang="en">
  <body>
    <p>
      <span id="c01_s001">Call me Ishmael. </span>
      <span id="c01_s002">Some years ago...</span>
    </p>
  </body>
</html>
\end{lstlisting}

\subsection{The Media Overlay Document (SMIL)}
%The synchronization logic is encapsulated in SMIL 3.0 XML files. 
A SMIL file acts as a temporal map that binds a specific XHTML content document to a corresponding audio file. For each XHTML chapter, a corresponding SMIL file is generated mapping the structural IDs ($id$) to the calculated audio time intervals $[t_{\text{start}}, t_{\text{end}}]$. 
The references are constructed using relative paths to ensure compatibility across reading systems. 
The \texttt{<seq>} element points to the target XHTML file via the \texttt{epub:textref} attribute, and \texttt{<par>} elements group the text anchor with the audio playback window:
\begin{lstlisting}[language=xml]
<smil xmlns="http://www.w3.org/ns/SMIL" version="3.0"
      xmlns:epub="http://www.idpf.org/2007/ops">
  <body>
    <seq id="seq1" epub:textref="../text/chapter1.xhtml">
      <par id="par_c01_s001">
        <text src="../text/chapter1.xhtml#c01_s001"/>
        <audio src="../audio/chapter1.mp3" 
               clipBegin="0:00:00.000" 
               clipEnd="0:00:02.540"/>
      </par>
    </seq>
  </body>
</smil>
\end{lstlisting}

\subsection{The Package Document (OPF)}
Finally, the OPF manifest binds these files together. The XHTML item entry includes a \texttt{media-overlay} attribute that points to the ID of the SMIL item. The OPF informs the reading system that when the user opens \texttt{chapter1.xhtml}, it should simultaneously load \texttt{chapter1.smil} for playback:
\begin{lstlisting}[language=xml]
<package ... version="3.0">
  <metadata>
    <meta property="media:duration" 
    refines="#smil_01">0:00:15.0</meta>
  </metadata>
  <manifest>
    <item id="text_01" href="text/chapter1.xhtml" 
          media-type="application/xhtml+xml" 
          media-overlay="smil_01"/> 
          
    <item id="smil_01" href="smil/chapter1.smil" 
          media-type="application/smil+xml"/>
          
    <item id="audio_01" href="audio/chapter1.mp3" 
          media-type="audio/mpeg"/>
  </manifest>
</package>
\end{lstlisting}

\section{Neural Text-to-Speech Engines} 
\label{sec:TTS}

Our framework supports two distinct state-of-the-art neural architectures for waveform synthesis, namely XTTS-v2 and Chatterbox.

\subsection{XTTS-v2 (Coqui):} XTTS-v2 is an autoregressive transformer model based on a GPT-2 architecture. It predicts discrete audio tokens conditioned on input text and a reference speaker embedding. These tokens are converted into a continuous waveform using a HiFi-GAN-based decoder adapted for XTTS \cite{kim2021conditional}. As an autoregressive model, it generates audio sequentially, enabling state-of-the-art zero-shot voice cloning and cross-lingual prosody transfer. However, this sequential dependency limits parallelization, resulting in higher computational latency.

\subsection{Chatterbox (Resemble AI):} Chatterbox employs a non-autoregressive Flow Matching architecture built on a 0.5B-parameter Llama transformer backbone. Unlike diffusion models that iteratively denoise signals, Flow Matching learns a continuous probability flow and solves it using an Ordinary Differential Equation (ODE) solver \cite{lipman2023flow}. This design enables stable, high-fidelity synthesis with fewer inference steps compared to traditional diffusion approaches. %Our implementation leverages alignment-informed inference for robust sentence boundary detection, reducing the hallucinations often observed in autoregressive systems.

\section{Methodology}
\label{secimp}

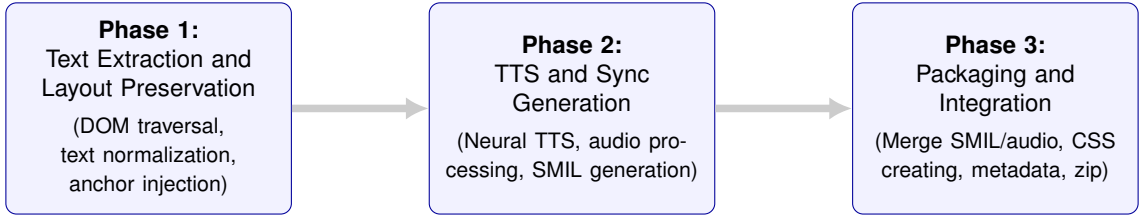
\begin{figure*}[htbp]
\centering
\begin{tikzpicture}[
    % INCREASED: Horizontal distance changed from 0.8cm to 1.8cm
    node distance=1.0cm and 1.8cm,
    phase_box/.style={
        rectangle,
        draw=blue!60!black,
        fill=blue!5,
        rounded corners,
        text width=3.5cm,
        align=center,
        minimum height=2.8cm,
        font=\sffamily\footnotesize
    },
    process_arrow/.style={
        -{Latex[length=3mm, width=3mm]},
        line width=1mm,
        draw=gray!40
    }
]

% Phase 1 Node
\node (phase1) [phase_box] {
    \textbf{Phase 1:}\\
    Text Extraction and\\
    Layout Preservation\\
    \vspace{0.1cm}
    \scriptsize
    (DOM traversal, text normalization, anchor injection)
};

% Phase 2 Node
\node (phase2) [phase_box, right=of phase1] {
    \textbf{Phase 2:}\\
    TTS and Sync\\
    Generation\\
    \vspace{0.1cm}
    \scriptsize
    (Neural TTS, audio processing, SMIL generation)
};

% Phase 3 Node
\node (phase3) [phase_box, right=of phase2] {
    \textbf{Phase 3:}\\
    Packaging and\\
    Integration\\
    \vspace{0.1cm}
    \scriptsize
    (Merge SMIL/audio, CSS creating, metadata, zip)
};

% Arrows
\draw[process_arrow] (phase1) -- (phase2);
\draw[process_arrow] (phase2) -- (phase3);

% Title (Optional)
\node (title) [above=0.2cm of phase2, font=\sffamily\bfseries] {EPUB to EPUB 3 Media Overlay using TTS};

\end{tikzpicture}
\caption{Overview of the three phases of the Calliope methodology.}
\label{fig:3}
\end{figure*}

In this section, we describe the proposed framework for converting a static EPUB file into an EPUB 3 with Media Overlays using Neural TTS. The pipeline consists of three distinct phases: (1) text extraction and structural preservation of the original source, (2) TTS synthesis, audio processing, and temporal synchronization (SMIL), and (3) packaging of the final EPUB 3 container. Figure \ref{fig:3} provides a short overview of the three phases.

\subsection{Phase 1: Text Extraction and Layout Preservation}
Let $D$ represent the Document Object Model (DOM) of an XHTML chapter file. We traverse $D$ to identify the set of block-level elements $\mathcal{B} = \{b_1, b_2, \dots, b_n\}$ (e.g., paragraphs, headers) that act as leaf nodes containing narrative text.

For each block $b_j$, the raw text content is extracted and normalized to resolve Unicode canonicalization and standardize punctuation. This normalized text is then segmented into an ordered sequence of sentences $S_j = \{s_{j,1}, s_{j,2}, \dots, s_{j,N_j}\}$ using a standard sentence tokenizer.

This segmentation $S_j$ serves a dual purpose: it defines the input text segments for the TTS engine in Step 2, and simultaneously establishes the structural anchors required for synchronized text highlighting. For each sentence, a structural anchor is injected into the DOM in the form of a \texttt{span} element with a globally unique identifier, $id_{j,k}$
\begin{equation}
    b_j' = \bigcup_{k=1}^{N_j} \langle \text{span id}=id_{j,k} \rangle s_{j,k} \langle / \text{span} \rangle
\end{equation}
These insertions occur strictly within the blocks $b_j$, preserving the parent element's CSS attributes while creating the precise granularity required for sentence-level highlighting.

\subsection{Phase 2: TTS and Audio-Text Synchronization}
\label{sec:step2}
Let $\mathcal{M}$ denote a neural TTS model. For each sentence $s_{j,k}$, the model generates a discrete audio waveform $w_{j,k}$
\begin{equation}
    w_{j,k} = \mathcal{M}(s_{j,k}, \theta_{voice})
\end{equation}
where $\theta_{voice}$ represents the reference speaker embedding, typically computed from a short audio sample. Neural TTS models operate under strict context window constraints; a failure occurs if either the input text sequence or the generated latent sequence exceeds the model's capacity. To guarantee execution stability, we implement a two-tier handling strategy:

\paragraph{1. Pre-emptive Heuristic (Input Constraint):} 
Let $|s_{j,k}|$ denote the character length of sentence $s_{j,k}$. If $|s_{j,k}|$ exceeds a conservative safety threshold $\lambda^+$ (e.g., 200 characters), the text is recursively split at the nearest whitespace median. Additionally, certain diffusion-based architectures, such as Chatterbox~\cite{chatterboxtts2025}, may degrade on extremely short inputs. To mitigate this, adjacent sentences are merged if $|s_{j,k}|$ falls below a minimum threshold $\lambda^-$ (e.g., 60 characters).

\paragraph{2. Reactive Error Handling (Output Constraint):} 
Certain text segments possess a high phonetic density relative to their character count. For instance, numerical years (e.g., "1997") expand significantly in token space, potentially causing the decoder to exceed its maximum limit even when the character count is low. Predicting the exact token yield of a text segment is non-trivial. To address this, the synthesis function $\mathcal{M}(s_{j,k})$ is wrapped in a localized exception handler. If the model triggers a token-overflow runtime error, the system catches the exception and triggers a recursive split of $s_{j,k}$ into $s_{j,k, \text{left}}$ and $s_{j,k, \text{right}}$. The resulting outputs are concatenated to form the complete waveform
\begin{equation}
    w_{j,k} = \mathcal{M}(s_{j,k, \text{left}}) \parallel \mathcal{M}(s_{j,k, \text{right}})
\end{equation}
where $\parallel$ denotes signal concatenation.

\subsubsection{Temporal Calculation and Synchronization}
\label{sec:temp}

Since the TTS engine processes each sentence independently, the resulting audio segments must be concatenated into a continuous chapter stream that maintains natural prosody without artifacts. Let $\tau(w)$ denote the duration of a waveform in seconds. To mitigate boundary artifacts (such as digital clicks or hallucinated breath sounds common in flow-matching models), a linear fade-out filter is applied to the terminal segment (e.g., 50 ms) of each waveform. Subsequently, to ensure distinct audio separation between sentences, a silence padding $\delta$ (typically 0.15 s) is appended.

The cumulative audio stream $A_j$ for a chapter $j$ is formed by the sequential concatenation of these processed waveforms
\begin{equation}
    \label{eq:audio_stream}
    A_j = (w_{j,1}' \parallel \delta) \parallel (w_{j,2}' \parallel \delta) \parallel \dots \parallel (w_{j,N_j}' \parallel \delta)
\end{equation}
where $w_{j,k}'$ represents the fade-out processed waveform.

The synchronization timestamps for the SMIL files are derived deterministically from the construction in Equation \eqref{eq:audio_stream}. Let $t_{j,k,\text{start}}$ and $t_{j,k,\text{end}}$ represent the clip begin and clip end times for sentence $s_{j,k}$. The start time is defined as the summation of all previous audio durations
\begin{equation}
    \label{eq:t_start}
    t_{j,k,\text{start}} = \sum_{i=1}^{k-1} (\tau(w_{j,i}') + \tau(\delta))
\end{equation}
To achieve a "gapless" visual experience, where the highlighting remains active during the silence $\delta$ between sentences, we extend the active interval of the current sentence to meet the start of the subsequent sentence
\begin{equation}
    \label{eq:t_end}
    t_{j,k,\text{end}} = t_{j,k+1,\text{start}}
\end{equation}
For the final sentence $N_j$, the end time is its natural duration $t_{j,N_j,\text{end}} = t_{j,N_j,\text{start}} + \tau(w_{j,N_j}') + \tau(\delta)$. This gapless continuity prevents the visual "flicker" effect often observed in audio-e-books where the highlight disappears during pauses. Furthermore, strict adherence to contiguous time intervals is critical, as certain reading systems validation checks will reject Media Overlays containing temporal gaps or overlaps.

\subsection{Phase 3: Packaging and Accessibility Integration}
The final phase involves reconstructing the EPUB container to comply with EPUB 3.3 Accessibility specifications.

\subsubsection{SMIL Generation}
For each XHTML chapter, a corresponding SMIL file is generated mapping the structural IDs ($id_{j,k}$) to the time intervals calculated in Step 2. References are constructed using relative paths to ensure compatibility across reading systems:
\begin{lstlisting}[language=xml]
<par id="par_f001">
    <text src="../text/chapter.xhtml#f001"/>
    <audio src="../audio/chapter.mp3" 
           clipBegin="t_start" clipEnd="t_end"/>
</par>
\end{lstlisting}

\subsubsection{Visual Accessibility (Dark Mode)}
To ensure the visual overlay remains legible across various reading environments, we inject an additional CSS rule utilizing the \texttt{prefers-color-scheme} media query. This dynamically adapts the highlight color (e.g., high-contrast yellow for Light Mode, muted purple for Dark Mode) based on the user's system preferences.

\subsubsection{Metadata and Packaging}
Finally, the Open Packaging Format (OPF) file is updated. We inject the \texttt{media:duration} metadata for every SMIL file and the total publication duration, calculated by summing the precise durations derived in Step 2. The \texttt{media:active-class} metadata is inserted to trigger the CSS highlighting mechanism. The resulting assets are zipped to produce the final validated EPUB 3 file with Media Overlays.

\section{Implementation and Usage}

The proposed framework is implemented as an open-source MIT Licenced Python library, designed for deployment on consumer-grade hardware (Linux/WSL, macOS, or Windows) with optional GPU acceleration.
%To ensure reproducibility and prevent dependency conflicts between the distinct neural synthesis engines, we recommend isolating the execution environments.

\subsection{Installation}
The repository supports the two TTS systemt \textit{Chatterbox}  and \textit{XTTS-v2}, see Section \ref{sec:TTS}.

\subsubsection{Environment Setup}
We recommend initializing a Python 3.11 virtual environment to manage the dependencies. This isolation is critical for handling the specific version constraints for \texttt{numpy} and \texttt{torch} mandated by the underlying audio processing libraries:
\begin{lstlisting}[language=bash]
# Clone the repository
git clone https://github.com/hugohammer/TTS-Narrated-Ebook-Creator.git
cd TTS-Narrated-Ebook-Creator
# Create and activate a virtual environment. It is 
# recommeneded to use separate environments for the 
# two TTS models.
python3.11 -m venv env_chatterbox
source env_chatterbox/bin/activate
python3.11 -m venv env_xtts
source env_xtts/bin/activate
\end{lstlisting}

\subsubsection{Dependencies}
The project relies on several core libraries: \texttt{EbookLib} for parsing the EPUB container, \texttt{BeautifulSoup4} for DOM manipulation, and \texttt{FFmpeg} for audio signal processing. The backend-specific dependencies can be installed via the provided requirement files:
\begin{lstlisting}[language=bash]
# For Chatterbox
pip install -r requirements/requirements_chatterbox.txt

# For XTTS-v2
pip install -r requirements/requirements_xtts.txt
\end{lstlisting}

\subsection{Execution}
The pipeline is executed via a Command Line Interface. The user must provide the source EPUB file and a reference audio sample (approximately 15 seconds of clean speech in WAV format) to define the narrator's voice profile.

\subsubsection{Running the Scripts}
To execute the creation process using the Chatterbox engine:
\begin{lstlisting}[language=bash]
python src/epub_processor_chatterbox.py input_book.epub \
       --voice assets/neutral_narrator.wav \
       --output output_book.epub
\end{lstlisting}
and for XTTS-v2
\begin{lstlisting}[language=bash]
python src/epub_processor_xtts.py input_book.epub \
       --voice assets/neutral_narrator.wav \
       --output output_book.epub
\end{lstlisting}

\subsubsection{Configuration Arguments}
The tool supports several runtime arguments to customize the output:
\begin{itemize}
    \item \verb|--gpu|: Forces the use of CUDA acceleration if available, significantly reducing synthesis time.
    \item \verb|--language [code]|: Specifies the target language for the TTS engine (default: 'en').
    \item \verb|--skip-audio|: Enables a debug mode that created the EPUB without running the computationally expensive TTS step.
\end{itemize}
Upon completion, the script outputs a validated EPUB 3 file containing the synchronized Media Overlays. The file can be uploaded to e.g. the Thorium Reader for Windows, MacOS and Linux or the BookFusion or Storyteller apps for Android/iOS.

\section{Evaluation of Alignment Strategies}

As discussed, a benefit of our implementation was that it ensured perfect synchronization between text highlighting and audio. However, from an implementation point of view, it could be argued that using existing forced alignment methods would have been easier, since TTS and alignment could have been separated during the creation of the narrated e-book. To analyse the potential of using forced alignment, we considered two recent methods: Afaligner~\cite{afaligner2025} and Storyteller~\cite{storyteller}. Afaligner was the forced alignment method used in Syncabook~\cite{syncabook}, was inspired by Aeneas, and was based on the Dynamic Time Warping algorithm~\cite{sakoe1978dynamic}. Storyteller employed a transcription-based alignment strategy. First, the audio was transcribed into time-stamped text using Whisper~\cite{radford2023robust}. This transcription was then mapped to the e-book's content using fuzzy string matching to resolve discrepancies between what was narrated in the audio and the source text.
\begin{figure*}[!htbp]
\centering
\includegraphics[width=\textwidth]{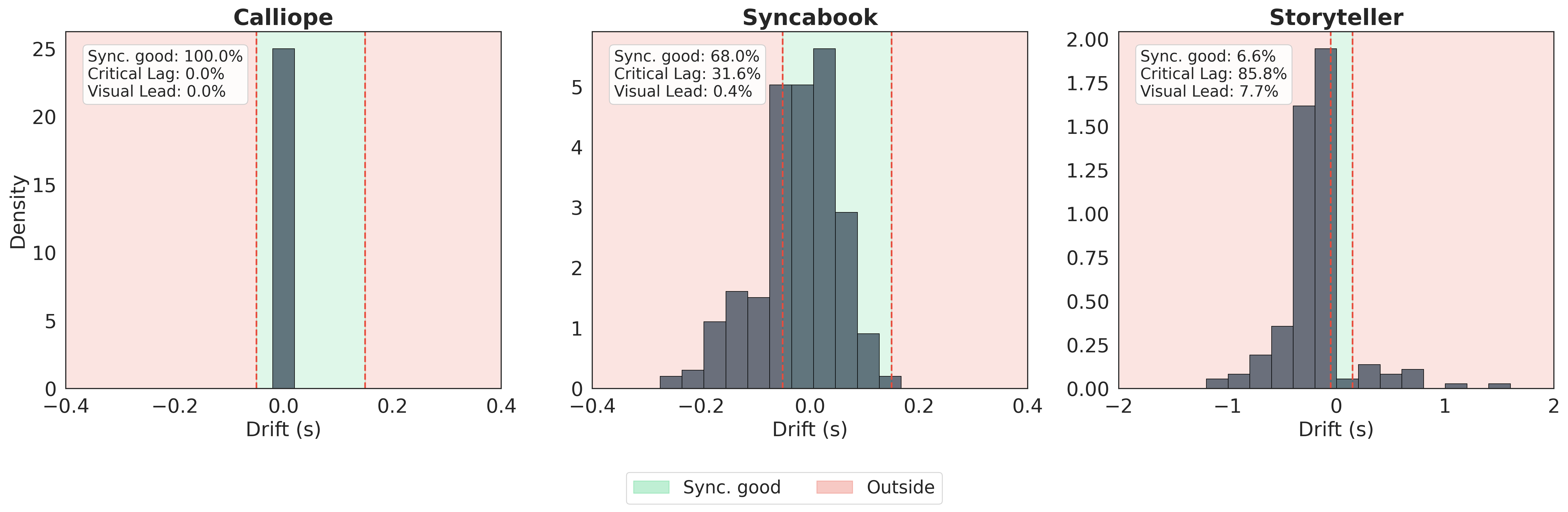}
\caption{Histograms of the drift distributions for the forced alignment methods. Note that the scale on the $x$ axis is different for the different figures. The green and red areas in the figures, shows where the drift is acceptable or where it may affect the reading experience. Our method is exact, and the drift was zero.
}
\label{fig:2}
\end{figure*}

The performance of the alignment methods was evaluated on the e-book \textit{The Gift of the Magi}, freely available from the Gutenberg Project~\cite{ohenry_magi}. The forced alignment methods received as input the e-book and the TTS audio obtained from Chatterbox and produced as output an EPUB 3 Media Overlay. The .epub file was further unzipped, and the time stamps of the SMIL files were compared against the exact time stamps obtained using our method described in Section~\ref{sec:step2}. The Storyteller method did not use the same sentence tokenizer as our method, and therefore some sentences were potentially different. We employed a filtering strategy where the time stamps were only compared for sentences that were identical and in the same chapter (85\% of the sentences).
\begin{table}
\centering
\caption{Key statistics for the synchronization drift distribution in seconds for our method and the forced alignment methods used in Syncabook and Storyteller. P10 and P90 refer to the 10\% and 90\% quantiles of the drift distributions, respectively.}
\label{tab:1}
\begin{tabular}{lcccccc}
\hline
\textbf{Method} & \textbf{Min} & \textbf{P10} & \textbf{Mean} & \textbf{Median} & \textbf{P90} & \textbf{Max} \\ \hline
Calliope & 0 & 0 & 0 & 0 & 0 & 0 \\
Storyteller & -34.91 & -0.75 & -1.11 & -0.20 & -0.01 & 1.60 \\
Syncabook & -0.96 & -0.14 & -0.03 & -0.01 & 0.07 & 0.49 \\
\hline
\end{tabular}
\end{table}

Table~\ref{tab:1} shows key statistics for the distribution of the drift for the different sentences for the two forced alignment methods. Figure~\ref{fig:2} shows histograms of the drift distributions. Please note that the scale on the x-axis is different for the three histograms. A negative value means that the text highlighting lagged behind the audio for the forced alignment method, while a positive value means that the text highlighting appeared before the audio. Readers are more sensitive to text highlighting lagging behind the audio than the other way around. Even a small lag of 50~ms could can affect a reading experience, while the text highlighting could typically be as much as 150~ms early before affecting the reading experience~\cite{conrey2003audiovisual,grant2001speech}. The green and red areas in the figure show the intervals where the drift was within and outside the acceptable interval, $[-50, 150]$ ms, respectively.

In Table~\ref{tab:1}, we see that the Storyteller forced alignment ran into substantial synchronization issues for some sentences, with drift of over 30 seconds. Syncabook was fairly consistent, with a maximal drift of 0.96 seconds. However, in the middle panel of Figure~\ref{fig:2}, we see that for over 30\% of the sentences, the lag was more than the delay of 50~ms and can affect the reading experience. For Storyteller, the synchronization was satisfactory for only 6.6\% of the sentences. Since our method does not rely on forced alignment, the drift was zero as shown in the upper row of Table~\ref{tab:1} and the left panel of Figure~\ref{fig:2}. 

The main takeaway is that even when the forced alignment algorithm worked as expected (Syncabook), the resulting drift could still represent an issue for the reader. This motivates the use of the exact alignment strategy developed in this paper.

\section{Conclusion}

In this work, we have presented a robust, open-source framework using SOTA open source TTS to convert standard text-based e-books into high quality narrated e-books in the EPUB 3 Media Overlays format. To the best of our knowledge this is the first open source solution to perform this task. Our solution offers several innovative steps. First, by using the exact timings from the TTS, resulting in perfect synchronization between text highlighting and audio. Our experiments showed that alternatively using forced alignment resulted in drift between audio and text significant enough to degrade the reading experience. Second, the surgical insertions strategy ensures that the original publisher's typography, layout, and embedded media were preserved. Finally, the implementation operates entirely offline, utilizing the local neural TTS models Chatterbox and XTTS-v2. This approach not only eliminates the costs of commercial APIs but also resolves privacy concerns and copyright compliance issues associated with uploading protected content to third-party cloud services. %Finally, the integration of robust safety protocols, including recursive text segmentation and dynamic visual accessibility for dark mode, ensures that the resulting publications meet modern accessibility standards (WCAG/EPUB Accessibility 1.1) and provide a seamless user experience across diverse reading systems.

% This research democratizes the production of high-quality, accessible digital literature. By providing a free, privacy-first toolchain, we empower libraries, archivists, and independent authors to upgrade their existing digital collections, transforming static text into immersive, multimodal experiences that are accessible to all readers, regardless of visual or cognitive ability.

A first natural step for future research is to add a graphical user interface, for the installation and the usage, to make the method easier accessible for more users. It is also interesting to explore quantized TTS models to evaluate the potential for using the method directly on mobile devices. 

\bibliographystyle{plain} % or 'abbrv', 'unsrt', 'plain'
\bibliography{bibl}

\begin{thebibliography}{10}

\bibitem{epub2tts}
aedocw.
\newblock {epub2tts}: Turn an epub or text file into an audiobook.
\newblock \url{https://github.com/aedocw/epub2tts}, 2024.
\newblock GitHub Repository.

\bibitem{casanova2024xtts}
Edresson Casanova, Kelly Davis, Eren G{\"o}lge, G{\"o}rkem G{\"o}knar, Iulian Gulea, Logan Hart, Aya Aljafari, Joshua Meyer, Reuben Morais, Samuel Olayemi, et~al.
\newblock Xtts: a massively multilingual zero-shot text-to-speech model.
\newblock {\em arXiv preprint arXiv:2406.04904}, 2024.

\bibitem{conrey2003audiovisual}
Brianna~L Conrey and David~B Pisoni.
\newblock Audiovisual asynchrony detection for speech and nonspeech signals.
\newblock In {\em AVSP}, volume 2003, pages 25--30, 2003.

\bibitem{daisystandard}
{DAISY Consortium}.
\newblock Daisy standard overview.
\newblock \url{https://daisy.org/activities/standards/daisy/}, 2025.

\bibitem{thorium2025}
{EDRLab}.
\newblock {Thorium Reader}.
\newblock \url{https://www.edrlab.org/software/thorium-reader/}, 2025.

\bibitem{elevenreader}
{ElevenLabs}.
\newblock Elevenreader official website.
\newblock \url{https://elevenreader.io/}, 2025.

\bibitem{audibleepub3maker}
Funway.
\newblock {audible-epub3-maker}: Generate audiobooks from plain epub files.
\newblock \url{https://github.com/funway/audible-epub3-maker}, 2025.
\newblock GitHub Repository.

\bibitem{grant2001speech}
Ken~W Grant and Steven Greenberg.
\newblock Speech intelligibility derived from asynchronous processing of auditory-visual information.
\newblock In {\em AVSP}, volume 200, pages 132--137, 2001.

\bibitem{jung2025impact}
Jookyoung Jung and Wenrui Zhang.
\newblock The impact of text-audio synchronized enhancement on collocation learning from reading-while-listening: an extended replication of.
\newblock {\em International Review of Applied Linguistics in Language Teaching}, 63(3):2201--2229, 2025.

\bibitem{kim2021conditional}
Jaehyeon Kim, Jungil Kong, and Juhee Son.
\newblock Conditional variational autoencoder with adversarial learning for end-to-end text-to-speech.
\newblock In {\em International Conference on Machine Learning}, pages 5530--5540. PMLR, 2021.

\bibitem{lipman2023flow}
Yaron Lipman, Ricky~TQ Chen, Heli Ben-Hamu, Maximilian Nickel, and Matt Le.
\newblock Flow matching for generative modeling.
\newblock In {\em 11th International Conference on Learning Representations, ICLR 2023}, 2023.

\bibitem{webspeechapi}
{Mozilla Developer Network}.
\newblock Web speech api documentation.
\newblock \url{https://developer.mozilla.org/en-US/docs/Web/API/Web_Speech_API/Using_the_Web_Speech_API}, 2025.

\bibitem{naturalreader}
{NaturalSoft Ltd}.
\newblock Naturalreader official website.
\newblock \url{https://www.naturalreaders.com/}, 2025.

\bibitem{ohenry_magi}
{O. Henry}.
\newblock {The Gift of the Magi}.
\newblock Project Gutenberg, January 2005.
\newblock eBook \#7256. Updated: Dec 24, 2021.

\bibitem{Pettarin_aeneas_2016}
Alberto Pettarin.
\newblock {aeneas}: A python/c library and set of tools to synchronize audio and text.
\newblock \url{https://github.com/readbeyond/aeneas}, 2016.

\bibitem{syncabook}
r4victor.
\newblock {syncabook}: A tool for creating ebooks with synchronized text and audio.
\newblock \url{https://github.com/r4victor/syncabook}, 2024.
\newblock GitHub Repository.

\bibitem{afaligner2025}
{r4victor}.
\newblock {Afaligner: A tool for forced alignment}.
\newblock \url{https://github.com/r4victor/afaligner/}, 2025.

\bibitem{radford2023robust}
Alec Radford, Jong~Wook Kim, Tao Xu, Greg Brockman, Christine McLeavey, and Ilya Sutskever.
\newblock Robust speech recognition via large-scale weak supervision.
\newblock In {\em International conference on machine learning}, pages 28492--28518. PMLR, 2023.

\bibitem{chatterboxtts2025}
{Resemble AI}.
\newblock {Chatterbox-TTS}.
\newblock \url{https://github.com/resemble-ai/chatterbox}, 2025.
\newblock GitHub repository.

\bibitem{sakoe1978dynamic}
Hiroaki Sakoe and Seibi Chiba.
\newblock Dynamic programming algorithm optimization for spoken word recognition.
\newblock {\em IEEE transactions on acoustics, speech, and signal processing}, 26(1):43--49, 1978.

\bibitem{storyteller}
Smoores.
\newblock Storyteller: Self-hosted platform for syncing audiobooks and ebooks.
\newblock \url{https://gitlab.com/storyteller-platform/storyteller}, 2025.
\newblock GitLab Repository.

\bibitem{epub3spec}
{World Wide Web Consortium}.
\newblock Epub 3.3 specification.
\newblock \url{https://www.w3.org/TR/epub-33/}, 2025.

\end{thebibliography}

\end{document}